\PassOptionsToPackage{dvipsnames}{xcolor}
\documentclass[sigplan,nonacm]{acmart}

\usepackage{xspace}
\usepackage{xfp} 
\usepackage{multirow}
\usepackage{multicol}
\usepackage{pifont}
\usepackage[dvipsnames]{xcolor}
\usepackage{color}
\usepackage{colortbl}
\usepackage{enumitem}
\usepackage{mdframed}
\usepackage{fvextra}
\usepackage[frozencache]{minted}
\setminted{
  fontsize=\footnotesize,
  breaklines=true,
  breakanywhere=true,
  frame=none,
  xleftmargin=0pt,
}
\definecolor{mintedhl}{HTML}{FBE0E0} 
\definecolor{mintedgreen}{HTML}{DFF5E1} 
\usepackage{subcaption}
\usepackage{pifont}

\definecolor{fragone}{RGB}{222,242,252} 
\definecolor{fragtwo}{RGB}{255,255,231} 
\newlength{\hlrule}
\newlength{\hlpad}
\makeatletter
\newlength{\hl@W}
\newcommand{\hl@cell}[2]{
  \setlength{\fboxsep}{0pt}%
  \hbox{\colorbox{#1}{\begin{minipage}[t]{\hl@W}\kern1pt #2\kern1pt\end{minipage}}}}
\newcommand{\hlband}[2]{
  \setlength{\hl@W}{\linewidth}\noindent\hl@cell{#1}{#2}}
\newcommand{\hlregion}[2]{
  \setlength{\hl@W}{\linewidth}%
  \noindent\vbox{\offinterlineskip
    \hl@cell{fragone}{#1}%
    \hl@cell{fragtwo}{#2}}}
\makeatother

\AtBeginDocument{%
  }

\setcopyright{acmlicensed}
\copyrightyear{2018}
\acmYear{2018}
\acmDOI{XXXXXXX.XXXXXXX}
\acmConference[Conference acronym 'XX]{Make sure to enter the correct
  conference title from your rights confirmation email}{June 03--05,
  2018}{Woodstock, NY}
\acmISBN{978-1-4503-XXXX-X/2018/06}

\renewcommand\footnotetextcopyrightpermission[1]{}

\newcommand{\para}[1]{\smallskip\noindent {\bf #1} }

\usepackage{tikz}
\newcommand*\circled[1]{\tikz[baseline=(char.base)]{
            \node[shape=circle,draw,inner sep=0.5pt] (char) {#1};}}

\newcommand{\specula}{Specula\xspace}

\newcommand{\tla}{TLA$^+$\xspace}

\newcommand{\TODO}[1]{\textcolor{red}{\textbf{TODO:} #1}}

\newcommand{\toreview}[1]{\textcolor{blue}{#1}}

\newcommand{\draftnum}[1]{#1}

\newcommand{\numcases}{\draftnum{48}}    
\newcommand{\numdistributed}{\draftnum{36}} 
\newcommand{\numconcurrency}{\draftnum{12}} 
\newcommand{\locmin}{\draftnum{2K}}      
\newcommand{\locmax}{\draftnum{95K}}     

\newcommand{\numbugs}{\draftnum{249}}    
\newcommand{\numnewbugs}{\draftnum{207}} 
\newcommand{\numconfirmedraw}{68}        
\newcommand{\numknownraw}{42}            
\newcommand{\numconfirmed}{\draftnum{\numconfirmedraw}}
\newcommand{\numknownbugs}{\draftnum{\numknownraw}}

\newcommand{\numfixed}{\draftnum{24}}    
\newcommand{\numreported}{\draftnum{89}} 

\newcommand{\mcbugs}{\draftnum{200}}     
\newcommand{\bfsbugs}{\draftnum{187}}    
\newcommand{\simbugs}{\draftnum{13}}     
\newcommand{\cexmedian}{\draftnum{9}}    

\newcommand{\pctsafety}{\draftnum{99.10\%}}  
\newcommand{\pctliveness}{\draftnum{0.90\%}} 
\newcommand{\pctprotocol}{\draftnum{21.1\%}}  
\newcommand{\pctcodelevel}{\draftnum{78.9\%}} 
\newcommand{\srccode}{\draftnum{87.35\%}}    
\newcommand{\srcissues}{\draftnum{74.34\%}}  
\newcommand{\srcfixes}{\draftnum{42.88\%}}   
\newcommand{\srcdocs}{\draftnum{20.04\%}}    

\newcommand{\gtbugs}{\draftnum{62}}      
\newcommand{\gtconfirmed}{\draftnum{34}} 
\newcommand{\gtbasic}{\draftnum{2}}      
\newcommand{\gttla}{\draftnum{3}}        
\newcommand{\recallbasic}{\draftnum{3.2\%}} 
\newcommand{\recalltla}{\draftnum{4.8\%}}   

\newcommand{\runhoursmedian}{\draftnum{3.69}} 
\newcommand{\runhoursmin}{\draftnum{1.43}}    
\newcommand{\runhoursmax}{\draftnum{9.86}}    
\newcommand{\runcostmedian}{\draftnum{\$57}}  
\newcommand{\runcostmin}{\draftnum{\$19}}     
\newcommand{\runcostmax}{\draftnum{\$168}}    
\newcommand{\costopus}{\draftnum{\$50.73}}       
\newcommand{\minsopus}{\draftnum{166}}           
\newcommand{\minssonnet}{\draftnum{188}}         

\newcommand{\numinvrevised}{\draftnum{36}}   
\newcommand{\numinstrfixed}{\draftnum{33}}   
\newcommand{\pcterrinv}{\draftnum{17.3\%}}    
\newcommand{\pcterrmodel}{\draftnum{60.5\%}}  
\newcommand{\pcterrinstr}{\draftnum{22.2\%}}  
\newcommand{\pctconfrepro}{\draftnum{47.5\%}}      
\newcommand{\pctconfdischarged}{\draftnum{48.8\%}} 
\newcommand{\pctconfnorepro}{\draftnum{1.0\%}}     
\newcommand{\pctonefix}{\draftnum{91.3\%}}     

\newcommand{\gtsonnet}{\draftnum{10}}          
\newcommand{\budgetsonnet}{\draftnum{61\%}}    
\newcommand{\costpergtsonnet}{\draftnum{\$59}} 
\newcommand{\costpergtopus}{\draftnum{\$16}}   
\newcommand{\fpcandsonnet}{\draftnum{39}}      
\newcommand{\fpcleansonnet}{\draftnum{33}}     

\begin{document}

\title{\specula: Scaling formal specifications for autonomous model checking 
  of system code}

\author{Qian Cheng}
\affiliation{%
  \institution{Nanjing University}
  \city{Nanjing}
  \state{}
  \country{China}
}

\author{Saad Mohammad Rafid Pial}
\affiliation{%
  \institution{University of Illinois}
  \city{Urbana-Champaign}
  \state{IL}
  \country{USA}
}

\author{Ruize Tang}
\affiliation{%
  \institution{Microsoft Research Asia}
  \city{Beijing}
  \state{}
  \country{China}
}

\author{Yiming Su}
\affiliation{%
  \institution{University of Illinois}
  \city{Urbana-Champaign}
  \state{IL}
  \country{USA}
}

\author{Emilie Ma}
\affiliation{%
  \institution{University of British Columbia}
  \city{Vancouver}
  \state{}
  \country{Canada}
}

\author{Finn Hackett}
\affiliation{%
  \institution{University of British Columbia}
  \city{Vancouver}
  \state{}
  \country{Canada}
}

\author{Ivan Beschastnikh}
\affiliation{%
  \institution{University of British Columbia}
  \city{Vancouver}
  \state{}
  \country{Canada}
}

\author{Yu Huang}
\affiliation{%
  \institution{Nanjing University}
  \city{Nanjing}
  \state{}
  \country{China}
}

\author{Tianyin Xu}
\affiliation{%
  \institution{University of Illinois}
  \city{Urbana-Champaign}
  \state{IL}
  \country{USA}
}

\begin{abstract}
\specula is a push-button agentic system that generates high-quality 
    formal specifications for large, complex system code and uses 
    these specifications for highly effective model checking and bug finding.
\specula employs large language model (LLM) based coding agents to autonomously
    develop \tla{} specifications, including invariants that describe
    correctness properties of the target system 
    and formal models that describe the system implementation at the right level of abstraction.
\specula operates fully autonomously, removing the high manual overhead traditionally required to apply formal methods to real-world systems.
To overcome common LLM limitations
    like reward hacking and hallucinations, \specula uses self-evolving loops
    that iteratively improve specification quality by enabling the agents to deepen
    their understanding of a system's code and its behaviors.
We have used \specula to check \numcases{} open-source systems.
\specula found \numbugs{} bugs including many deep bugs that are hard to find with existing approaches.
\specula has been used by several companies and is open-sourced
    at \url{https://github.com/specula-org/Specula}.
\end{abstract}

%
%

\renewcommand\footnotetextcopyrightpermission[1]{}
\maketitle

\section{Introduction}

Formal specifications are key to automated reasoning about the correctness
    of large, complex software systems and 
    are developed in practice to check and verify the designs
    of important systems~\cite{howard2025ccf,bornholt2021s3,cauli2026,newcombe2015tlaaws,Li:nsdi:2026}.
Recently, formal specifications are further applied to {\it system implementations},
    serving as abstractions of system code~\cite{ouyang2025remix,Converos}.
These {\it system specifications}, empowered by model checking and theorem proving,
    help find deep bugs that manifest only under specific 
    thread interleavings, message orderings, and/or partial failures.

Despite their utility, formal specifications are historically considered expensive
    to develop and maintain, especially for real-world systems with evolving codebases.
In our prior work on specifying ZooKeeper and Asterinas~\cite{ouyang2025remix,Converos},
    writing high-quality specifications in the \tla{} specification language
    took months of effort, as it demanded
    expertise in system design and implementation,
    maintenance of conformance between code and specification,
    and mastery of the \tla{} language and toolchains.
As a result, our efforts were limited to specific system projects
    and are hard to scale.





Recent advances in large language models (LLMs) and agentic AI,
    embodied by coding agents like Claude Code and Codex,
    show promise in scaling formal specifications to
    {\it any} real-world systems, enabling
    {\it autonomous} formal reasoning with minimal human intervention.
Several recent studies reported that LLMs
    can generate a variety of formal specifications~\cite{ma2024specgen,wen2024enchanting,yang2025sespec,cheng2026sysmobench}.
However, directly using LLMs or agents to generate 
    formal specifications is fundamentally limited.
First, LLMs make mistakes and can generate imprecise and even hallucinated specifications~\cite{cheng2026sysmobenchsigops}.
Adding human-in-the-loop to review auto-generated specifications is not only expensive
    but also unreliable.
Moreover, as LLMs are known for reward hacking~\cite{gao2023overoptimization,pan2022misspecification}, 
    relying on LLMs to review and repair
    imprecise/incorrect specifications without well-defined boundaries is untenable.
Furthermore, an inherent challenge, no matter whether for human or AI, 
    is to decide the right level of abstractions to formally model complex systems~\cite{ouyang2025remix}.
These limitations are unlikely to disappear in next-generation LLMs or agents, as emergent model designs
    are not engineered to address them.




\para{Contributions.}
We present \specula, a push-button agentic system 
    that generates high-quality formal specifications for {\it any} large, complex system code and uses 
    the specifications for highly effective model checking and bug finding.
\specula employs LLM-based agents
    to autonomously develop \tla{} specifications, including 
    (1) invariants that describe correctness properties of the target system 
    and (2) formal models that describe the system implementation with the right level of abstractions.
Because distributed and concurrent systems form critical infrastructure requiring automated formal reasoning,
    \specula adopts \tla{} as its primary specification language.
\specula equips its agents with skills and tools and enables 
    them to use static analyzers,
    the TLC model checker~\cite{yu99tlc}, and trace validator (\S\ref{sec:implementation}).
\specula applies to system code written in any programming language as it
    abstracts the code in the \tla{} model.

\specula is {\it fully autonomous} in developing formal specifications
    and eliminates the barrier of applying formal methods to real-world system code.
\specula automatically checks the conformance between the formal \tla{} model 
    of the system and the system code,
    eliminating classic model-code gaps~\cite{amazon-use-tlaplus}.
Conformance checking is performed using automated
    trace validation~\cite{Converos,cirstea2024trace,hackett2025tracelink},
    where the agents autonomously instrument system code,
    collect traces, and validate generated \tla{} model
    using code-level traces.
If the \tla{} model fails trace validation, \specula autonomously
    repairs the model or the code instrumentation
    until model-code conformance.

\specula anticipates mistakes and errors, as well as reward-hacking behaviors, 
    introduced by LLMs 
    in generating specifications, including both system models 
    and invariants.
\specula crafts self-evolving loops where the agents iteratively improve specification quality. 
In ambiguous situations, e.g., an invariant violation can be caused by 
    incorrect modeling, bugs in the code, wrongly inferred invariants, 
    or their combinations, the evolving 
    loops enable agents to
    deepen their understanding of system code and its behaviors
    and pinpoint true root causes.
To prevent reward hacking (e.g., overfitting a system model to match code-level traces),
    \specula defines clear boundaries by pairing trace validation 
    and model checking---the former ensures that the model admits 
    code-level behaviors and the latter rejects invalid states.    

\specula models system implementations with the levels of abstractions 
    guided by the invariants
    it inferred from the system artifacts (e.g., code, comments, documents, and user-reported issues).
As a principle, it abstracts out behaviors irrelevant to the correctness properties
    while describing important details that must be carefully modeled. 
It thus can generate multiple system models with different abstractions 
    for different kinds of invariants, balancing efficiency (avoiding state-space explosion)
    and utility (bug finding).

With the specification in place, \specula 
    checks the formal model to detect invariant violations.
Each violation is a potential bug in the system's code.
\specula replays each violation at the code level using the model-level traces and
    encapsulates the violation in a system test to help developers reliably reproduce the bug.

\para{Key results.}
We maintain \specula as an open-source project.
\specula currently supports agents such as Claude Code, Codex, and Copilot CLI
    with different LLMs.
We have used \specula (with a default configuration of Claude Code with Opus-4.8) 
    to check \numcases{} open-source systems, including
    both distributed systems (e.g., MongoDB, Etcd, and ScyllaDB)
    and concurrent systems (e.g., GCC libgomp and LLVM libomp).
\specula found \numbugs{} bugs including many deep bugs that are hard to find by existing approaches.
\specula reports no false positive as all the bugs are reproduced at the code level.
We reported \numreported{} bugs;
    so far, \numconfirmed{} have been confirmed and \numfixed{} have been fixed. 
\specula has been used by several companies. 
The \specula project is maintained at:
\begin{center}
    \url{https://github.com/specula-org/Specula}.
\end{center}

\section{Background}
\label{sec:background}

\subsection{Formal specification and model checking}
\label{sec:background-tla}

\specula targets concurrent and distributed systems
    and thus chooses \tla{} as the specification language.
\tla{} is a language for writing formal models as abstractions
    of programs;
    it is not specific to any implementation languages like C/C++, Java, Rust, etc.
\tla{} has been widely used for checking system designs and distributed protocols
    in practice~\cite{newcombe2015tlaaws,howard2025ccf};
    recently, it has also been used for checking
    code-level behaviors by extracting models from code~\cite{ouyang2025remix,Converos,cheng2026sysmobenchsigops}.

A \tla{} specification comprises two parts:
    a \emph{model} that describes behaviors of the system
    and \emph{invariants} that define its correctness properties.
The model specifies system behaviors as a collection of state variables,
    an initial predicate that defines their initial values,
    and a next-state relation that determines state transitions.
The next-state relation is expressed as multiple \emph{actions},
    each describing an atomic state update of the variables.
An invariant is a property that must hold in every reachable state of the model.


\begin{figure}[t]
    \centering
    \begin{subfigure}{\linewidth}
\begin{minted}{cpp}
void fsm::append_entries(...) {
\end{minted}\par\nobreak
        \hlregion{
\begin{minted}{cpp}
%\circled{1}%  index_t last_new_idx = request.prev_log_idx;
    if (!request.entries.empty()) {
    last_new_idx =
        _log.maybe_append(std::move(request.entries)); }
\end{minted}}{
\begin{minted}{cpp}
%\circled{2}%  advance_commit_idx(
    std::min(request.leader_commit_idx, last_new_idx)); }
\end{minted}}
        \caption{Code snippet in C++}
        \label{fig:commit-code}
        \vspace{5pt}
    \end{subfigure}
    \begin{subfigure}{\linewidth}
        {\setminted{xleftmargin=\hlpad}\input{code/scylla-commit-clamp-1.tex}}\par\nobreak
        {\setminted{xleftmargin=\hlpad}\hlregion{\input{code/scylla-commit-clamp-2.tex}}{\input{code/scylla-commit-clamp-3.tex}}}
        \caption{The \tla{} action that models the code snippet in (\ref{fig:commit-code})}
        \label{fig:commit-tla}
        \vspace{5pt}
    \end{subfigure}
    \begin{subfigure}{\linewidth}
        \input{code/scylla-commit-inv.tex}
        \caption{A safety invariant of the above code snippet---a
            non-stale leader's log matches every entry each node has committed.}
        \label{fig:commit-inv}
    \end{subfigure}
    \caption{An example of \tla{} specification, including (\ref{fig:commit-tla})
        model and (\ref{fig:commit-inv}) invariant of (\ref{fig:commit-code}) 
        the source code snippet in C++ (from the Raft 
        implementation of ScyllaDB).}
    \label{fig:scylla-commit-transform}
    \vspace{-10pt}
\end{figure}

Figure~\ref{fig:commit-code} shows a source-code snippet in C++
    of the Raft implementation in ScyllaDB, which is a 
    part of the handler code a node runs upon receiving log entries from the leader.
The handler appends the entries to its local log \circled{1},
    and then advances its commit index, which is
    capped by \textsf{std::min} \circled{2} to stay within the entries it actually holds.
Figure~\ref{fig:commit-tla} shows the \tla{} model of the code snippet 
    as an action (generated by \specula).
It is enabled when an append-entries message is pending,
    and describes steps \circled{1} and \circled{2}
    that update the state variable \textsf{commitIndex}
    following the control flow of the code.
The invariant \textsf{CommitIndexSafety} defines the correctness property
    the model must maintain:
    log entries that any node has committed agree with a non-stale leader's log.

\tla{} specifications can be verified using explicit-state
    model checking via TLC~\cite{yu99tlc} and
    symbolic model checking via Apalache~\cite{konnov2019apalache}.
\specula uses TLC which systematically explores all reachable states of the model
    to ensure that the invariants hold over the entire state space, or up to some depth bound.
When an invariant is violated, TLC returns a \emph{counterexample},
    a sequence of actions from an initial state to the violating state.
The number of reachable states grows combinatorially with the number
    of state variables and the size of the model.
Hence, the cost of model checking, in terms of both time and memory,
    is a key consideration that drives specification design~\cite{ouyang2025remix}.

\begin{figure}[t]
    \centering
    \includegraphics[width=\columnwidth]{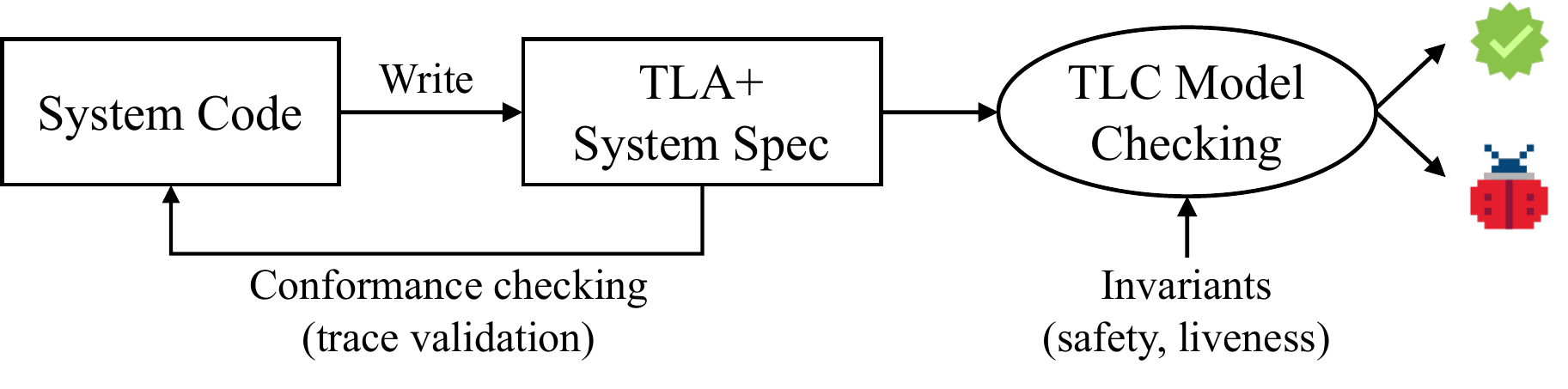}
     \vspace{-20pt}
    \caption{The classic process of formal specification and model checking; specifications 
        are written by human experts.}
    \vspace{-10pt}
    \label{fig:process}
\end{figure}

For \tla{} models that describe system implementations, 
    the model must conform to the code.
Conformance is commonly checked via 
    \emph{trace validation}~\cite{bornholt2021s3,cirstea2024trace,hackett2025tracelink,Converos}.
Trace validation checks whether execution traces collected from the running system code
    correspond to paths in the model's state space.
The traces are obtained by instrumenting the system code.
A trace the model rejects pinpoints where the model and the code diverge.
Figure~\ref{fig:process} shows the classic process of formal specification
    and model checking.

\subsection{AI for Formal Specification}
\label{sec:background-agents}

AI agents have proven effective at software engineering.
A coding agent can autonomously read large codebases,
    explain how code paths behave,
    and carry out tasks such as code generation and bug fixing~\cite{yang2024sweagent,chen2021codex,xia2023apr}.
These capabilities extend to formal specification:
    recent benchmarking finds that coding agents produce \tla{} models
    that pass syntax and runtime checks, capture essential system behavior,
    and reproduce known bugs in five real-world systems~\cite{cheng2026sysmobench}.

An increasing paradigm of agentic AI is to organize them as evolving loops,
    where the agent accumulates knowledge and improves its capabilities over iterations
    based on feedback from previous iterations~\cite{alphaevolve,openevolve}.
Evolving loops improve coding agents
    across tasks such as debugging and program repair~\cite{chen2024selfdebug,bouzenia2025repairagent}.
\specula organizes coding agents as self-evolving loops (\S\ref{sec:evo-loop}):
    each iteration presents an agent with new information,
    e.g., a counterexample, a model-code gap, a reproduction failure, etc.

\para{Limitations of AI agents.} Our experience tells that directly using 
    AI, even with the strongest LLMs, is fundamentally limited.
First, 
the codebase of large, complex systems often exceeds an agent's context size.
As a result, AI-generated \tla{} models often misrepresent the implementation:
    they hallucinate execution paths, oversimplify concurrency,
    and fallback on \tla{} models from their training data~\cite{cheng2026sysmobenchsigops}.
We found that AI often generates models that omit important behaviors 
    while admitting states the system never reaches.

Second, AI agents have no reliable basis
    to determine the right level of abstraction
    without expert guidance. 
Historically, human experts make empirical decisions based on intuition and experience~\cite{ouyang2025remix}.
Too high-level models are less useful (e.g., for finding deep bugs in code),
    while too low-level models include unnecessary details 
    that lead to state space explosion.
It is even harder for agents to discover invariants.
Real-world code rarely states them,
    and for well-known protocols agents often default to textbook invariants
    even in cases where the implementation deliberately deviates from such standards.

Lastly, reward hacking remains a severe problem of frontier AI.
When the agent is rewarded for a given goal (e.g., trace validation),
    it can overfit the specification by weakening invariants
    or manipulating \tla{} models only for the purpose of matching
    expected traces.
Such a behavior often results in imprecise models, leading to false results.
Unfortunately, reward hacking is often subtle and hard to catch.
\section{\specula Design}
\label{sec:specula}

\specula is a push-button agentic system that generates
    formal specifications for any given system
    and uses the specifications for model checking and bug finding.
The specification of a system generated by \specula includes
    (1) a formal model that effectively describes 
        the implementation of the target system
    and (2) invariants that describe the correctness properties
        the target system must hold.
\specula then uses the generated specifications to check
    the system implementation---it 
    systematically explores the formal model
    and checks for violations of invariants (which indicate bugs in the code).
For each violation, \specula further replays it 
    at the code level based on the model-level traces.

\specula autonomously generates formal specifications from the 
    artifacts of the target system (e.g., its repositories)
    using LLM-based coding agents such as Claude Code and Codex.
\specula aims to generate formal specifications that:

\begin{itemize}[leftmargin=*]
        \item precisely and comprehensively describe safety and liveness
        invariants the system must hold, including
        invariants at the implementation level that are not described
        in the protocol (if any);
        \item effectively model the system by describing important code-level behaviors
        that are critical to finding deep bugs while making the model
        tractable for model checking to avoid state-space explosion;
        \item conform to the system code so that model checking
            results are sound and can be reproduced at the code level.
\end{itemize}

\specula 
    iteratively improves 
    formal specifications until they meet the aforementioned criteria.
Our key insight is that, with the capabilities of frontier LLMs and agents (\S\ref{sec:background-agents}),
    generating formal specifications is no longer laborious and time-consuming as before (a coding agent 
    can generate quality \tla{} specifications in seconds~\cite{cheng2026sysmobench});
    however, generating precise, comprehensive, and effective formal specifications with no 
    model-code gap is still challenging and is beyond the raw capabilities of frontier LLMs/agents.
\specula is designed to provide agents with an effective process to guide 
    the generation of formal specifications at the right abstraction,
    and a rigorous harness to evolve the specifications to achieve
    model-code conformance and bug reproduction.

\specula targets concurrent and distributed systems
    which have complex, non-deterministic behaviors
    and require automated formal reasoning.
\specula uses \tla{} as the formal specification language and uses TLC~\cite{yu99tlc} 
    as the model checker,
    which are widely used for checking and verifying 
    concurrent and distributed systems~\cite{cheng2026sysmobench,ouyang2025remix,tang2024sandtable,Converos,cirstea2024trace}.
In principle, \specula could support any other specification language, but we have only
    experimented with \tla{}.

\subsection{Understanding correctness properties}
\label{sec:inv}

\specula starts from understanding the correct properties of the target system.
The correctness properties guide specification generation (\S\ref{sec:specgen})---the \tla{} model 
    of the system should preserve necessary details that can affect the properties
    and abstract out the others.
The correctness properties are embodied in {\it invariants} under {\it fault models}. 

\specula instructs AI agents to summarize two kinds of invariants from the system artifacts: 
    {\it protocol-level}
    invariants\footnote{We take a broad definition of {\it protocol} and assume each system is implemented
        based on an explicit or implicit protocol.} and {\it code-level} invariants,
    as shown in Figure~\ref{fig:invariants}.
\specula then encodes these invariants in \tla{}.

\begin{figure}[t]
    \centering
    \begin{subfigure}{\linewidth}
        \input{code/raft-committed-durable.tex}
        \caption{Protocol-level invariant generated by \specula for Etcd-Raft}
        \label{fig:inv-protocol}
        \vspace{5pt}
    \end{subfigure}
    \begin{subfigure}{\linewidth}
        \input{code/mongo-committed-in-log.tex}
        \caption{Code-level invariant generated by \specula for MongoDB}
        \label{fig:inv-code}
    \end{subfigure}
    \caption{Invariants summarized by \specula from artifacts}
    \label{fig:invariants}
\end{figure}

The protocol-level invariants are typically sourced 
    from protocol descriptions, system documentation,
    and source code.
For example, for a Raft implementation, protocol-level invariants often
    include Election Safety, Log Matching, Leader Completeness, etc.~\cite{ongaro2014raft}.
Figure~\ref{fig:inv-protocol} shows a protocol-level invariant \specula
    generated for Etcd-Raft, which states that 
    ``{\it committed entries are durable
    and will eventually be executed by all of the available state machines}~\cite{ongaro2014raft}.''
    Etcd-Raft's implementation follows this invariant.

\specula also infers code-level invariants from system code, test cases,
    issues, and revision histories. 
\specula equips the agent with a skill and few-shot examples that capture common patterns
    of code-level properties, such as atomicity of operations,
    bounded capacity of data structures, 
    and monotonicity of commit index;
it analyzes test cases to understand scenarios and the expected behavior;
it also reads scenarios described by user-reported issues and pull requests
    which often describe faults and the expected behavior.
Note that the agents are instructed to focus on summarizing 
    invariants on system-level behaviors; 
    they do not produce low-level invariants on APIs or variables (e.g., non-pointers).

Figure~\ref{fig:inv-code} shows a code-level invariant \specula generated
    for MongoDB whose replication follows the Raft protocol.
In this case, \specula finds that MongoDB does not strictly follow the Raft protocol 
    and thus the \textsf{CommittedInDurableStorage}
    invariant in Figure~\ref{fig:inv-protocol} does not apply.
To optimize write latency, MongoDB can be configured to report an entry as committed once a
    majority holds it in memory, before flushing it to stable storage. %
The design deliberately gives up the durability guarantee for a weaker one:
    a server never reports an entry as committed before that entry is written in its own log.
\specula derives this invariant from the revision history in which MongoDB began
    deciding commitment from the in-memory log rather than from stable
    storage~\cite{server85701}.

Each invariant is associated with a fault model, which guides
    the fault injection during model checking (\S\ref{sec:findbugs}).

\specula enforces the AI agents to provide concrete evidence of where
    it derived the invariants from (e.g., code, issues, commits, and comments).
The lineage enforces the agents to generate grounded invariants
    and is also used in the self-evolving loop that involves improving~invariants~(\S\ref{sec:evo-loop}).

\subsection{Generating effective system models}
\label{sec:specgen}

With correctness properties,
    \specula generates \tla{} models of the system.
An effective model is 
\emph{tractable} for model checking to explore its state space within a practical budget,
    and has \emph{high utility} for finding code bugs.
Historically, human experts made
    decisions on what to omit and what to model.
Recent work proposes to customize \tla{} models by specifying
    system modules at different abstraction levels~\cite{ouyang2025remix}.

\specula uses agentic AI to fundamentally reduce
    the cost of generating \tla{} models
    (see \S\ref{sec:background})
    and ensuring their code conformance (\S\ref{sec:conformance}).
With the low cost of generating high-quality models,
    \specula pushes the principle of {\it model customization} to a new level---it generates customized 
    models for important scenarios; these {\it scenario-based models}
    help model checking to focus on the system's
    most significant and potentially vulnerable behaviors, 
    increasing the likelihood of detecting bugs that may only manifest under specific conditions.
Concretely, \specula maintains a reference model of the target system (\S\ref{sec:base-model})
    and projects it to customized models based 
    on the scenarios it generates (\S\ref{sec:scenario-model}).

\subsubsection{Reference model.}
\label{sec:base-model}

\specula instructs AI agents to generate a reference \tla{} model 
    by reading and understanding the system code together with auxiliary materials
    in the artifact such as documents and test cases.
\specula lets AI agents autonomously determine the abstraction levels
    of different parts of system code based on the correctness properties (\S\ref{sec:inv})---the reference model 
    must describe all the behaviors relevant to each invariant
    and omit the others.
For example, data structures of hash implementations are typically abstracted out;
    however, for Papaya (a concurrent hash map), \specula explicitly models
    the per-slot meta byte and entry pointer
    to check the invariant that no entry is lost.
In contrast, the hash function itself stays out of the model, as no invariant depends on it.

Invariants guide the abstraction level,
but they do not determine the semantics of the selected behavior.
Once \specula models certain parts of the implementation,
    it derives the state representation and transition relation from the code.
Model variables correspond to implementation fields and data structures,
    and each action follows the guards, branches, and state updates at specific program points.

Figure~\ref{fig:scylla-commit-transform} shows how \specula
    models the append-entries handler of ScyllaDB-Raft
    (written in C++) using a \tla{} action, 
    guided by an invariant on the leader's \textsf{commitIndex}.
The resulting \tla{} action mirrors the original control flow.

The aforementioned generation alone does not ensure the code conformance of the reference model.
\specula ensures conformance by conformance checking and repair (\S\ref{sec:conformance}).


\subsubsection{Generating scenarios.}
\specula automatically generates scenarios from the system artifact,
    instead of relying on developers as in prior work~\cite{Li:nsdi:2026,ouyang2025remix,Converos}.
In our experience, state-of-the-art agents have strong capabilities of 
    summarizing a comprehensive set of scenarios that are described 
    in documents and comments, discussed in reported issues and revision history, 
    and encoded in test cases.
\specula enforces the AI agent to provide {\it evidence} when generating 
    a scenario---specific commits, issues, comments, 
    or other information that describes the scenario---making 
    the scenario auditable and minimizing hallucination.
Each scenario description also lists the related variables, actions,
    and invariants (which are used to generate scenario-based models).

\begin{figure}[t]
    \centering
\begin{minted}{yaml}
Scenario. Joint consensus quorum tracking on voter demotion:

Evidence:
  history: Three fixes reworked joint-config voter tracking
           (8f64a6d2d2, f31f73b1e8 #10618, b3cb4f3966)
  code: * broadcast_read_quorum (fsm.cc:1052) sends a read 
          quorum to followers whose cached vote flag is set
        * set_configuration (tracker.cc:101) sets that flag
          from the current config
        * committed<read_id> (tracker.cc:178) commits a read
          on acks from both current and previous voter sets

Modeling proposal:
  variables: * config[s]: (current, previous) configs
             * voters[s][c]: voters in config c
             * canVote[s]: cached send-path flag
  actions: EnterJoint, LeaveJoint, ReadBarrier, ...
  invariant: ReadBarrierProgress -- an issued read barrier
             eventually commits
\end{minted}
    \vspace{-10pt}
    \caption{A simplified modeling plan produced by \specula's analysis for ScyllaDB's Raft library.}
    \label{fig:scylla-brief}
\end{figure}

Figure~\ref{fig:scylla-brief} shows a scenario of joint-consensus reconfiguration during voter demotion
    in ScyllaDB, where a node is a voter in the previous configuration but a non-voter in the current one.
\specula creates this scenario because it observes that (1) the related code 
    has been changed by multiple fixes and (2) the fixes are inconsistent in different code snippets.
In fact, this scenario enabled \specula to discover a new bug that stalls a read barrier during 
    voter demotion, which violates the \textsf{ReadBarrierProgress} invariant.

\subsubsection{Scenario-based models.}
\label{sec:scenario-model}

For each scenario, \specula generates a \tla{} model as
    a {\it projection} of the reference model---it only describes behaviors 
    relevant to the target scenario, which is a subset of the behaviors the reference model 
    admits.
Each scenario-based model simplifies certain actions and behaviors in the reference model
    by rewriting the corresponding ones, while directly invoking the others;
    the reference model is immutable.
The projection is in principle sound: any invariant violation
    it finds is also a violation in the reference model.
The projection takes three steps (as exemplified in Figure~\ref{fig:etcd-projection}):
\begin{enumerate}[leftmargin=*]
\item {\bf Selecting actions.} \specula enables only actions needed to exercise the scenario
    and bounds the occurrence of those actions.
In Figure~\ref{fig:etcd-proj-select}, the scenario disables crashes (\textsf{CrashLimit = 0})
    and keeps lease reads (\textsf{ReadRequestLimit = 2}), focusing the search on the stale-leader read path.
    Setting an action's bound to 0 disables that action.

\item {\bf Coarsening actions.}
\specula can replace a multi-step process in the reference model 
    with a coarse action, if the behavior is required by the scenario 
    but not the targeted behavior.
Figure~\ref{fig:etcd-proj-coarsen} shows \textsf{MCLeaderElection} replacing the full voting protocol
    with one atomic action that picks a node whose log is up-to-date for a majority.
The reference model's election semantics are preserved (only valid leaders are chosen), 
    but the votes, timeouts, and re-candidacies are skipped.

\item {\bf Shaping action sequences.}
\specula can constrain an action to a particular phase of the execution, 
    forcing it to be atomic (if intermediate states are irrelevant)
    or otherwise encode the ordering assumptions of the target scenario.
Figure~\ref{fig:etcd-proj-shape} shows one such pattern.
\end{enumerate}

\begin{figure}[t]
    \centering
    \begin{subfigure}{\linewidth}
        \input{code/etcd-projection-step1.tex}
        \vspace{-7.5pt}
        \caption{{\bf Action selection.} 
            Wrapping an action with a counter that bounds its occurrences. 
            Setting the counter to 0 disables the action.}
        \label{fig:etcd-proj-select}
        \vspace{5pt}
    \end{subfigure}
    \begin{subfigure}{\linewidth}
        \input{code/etcd-projection-step2a.tex}
        \input{code/etcd-projection-step2.tex}
        \vspace{-7.5pt}
        \caption{{\bf Action coarsening.} 
            Replacing multiple actions with one atomic action. 
            Intermediate states are abstracted out.}
        \label{fig:etcd-proj-coarsen}
        \vspace{5pt}
    \end{subfigure}
    \begin{subfigure}{\linewidth}
        \input{code/etcd-projection-step3a.tex}
        \input{code/etcd-projection-step3.tex}
        \vspace{-7.5pt}
        \caption{{\bf Serialization.} 
            Serializing two actions into one atomic transition 
            via TLA+ action composition (\textsf{\textbackslash cdot}).}
        \label{fig:etcd-proj-shape}
    \end{subfigure}
    \caption{Three projection operations applied to Etcd-Raft's LeaseRead stale-read scenario.}
    \label{fig:etcd-projection}
    \vspace{-7.5pt}
\end{figure}

\subsection{Ensuring model-code conformance}
\label{sec:conformance}


\specula ensures model-code conformance of both reference and scenario-based models
    by trace validation~\cite{Converos,cirstea2024trace,ouyang2025remix}.
Historically, trace validation is costly due to {\it code instrumentation} (for producing code-level traces),
    which often requires manual effort~\cite{ouyang2025remix}.
\specula automates instrumentation as the model is generated from the code;
    it records the mapping from each action/variable in the model to
    the corresponding program location in the code during modeling (\S\ref{sec:specgen}).
\specula also repairs the models that fail conformance checking.

\subsubsection{Trace validation.}
\label{sec:trace_validation}
\specula implements automatic trace validation in three steps.
First, \specula uses AI agents
    to emit an instrumentation plan based on the input model.
For each \tla{} action, the plan lists the action's origin in the code, 
    a triggering point related to a designated operation, 
    and the state variables to capture.
Second, \specula uses the agent to instrument the system implementation at the 
    origin location with calls to a tracing library that records, for each action invocation, 
    the action name and the captured state.
Third, \specula instructs the AI agent to generate
    a \tla{} replay harness in which every model-level action 
    is wrapped to match the incoming event, execute the code-level action, 
    and check that the post-state agrees with the recorded snapshot.
TLC then checks the harness against collected traces, advancing one event at a time 
    until the trace is exhausted.
Any failures of trace validation mean that the model does not conform
    to the code and must be repaired.


\begin{figure}[t]
    \centering
    \begin{subfigure}{\linewidth}
        \input{code/repair-commit-broken.tex}
        \input{code/repair-commit-fixed.tex}
        \caption{The overfitted repair (red) and the correct repair (green)
            of the follower's accept path in Kudu-Raft.}
        \label{fig:repair-broken}
        \vspace{5pt}
    \end{subfigure}
    \begin{subfigure}{\linewidth}
        \input{code/repair-commit-inv.tex}
        \caption{Protocol-level invariant (State Machine Safety)}
        \label{fig:repair-inv}
    \end{subfigure}
    \caption{An overfitted repair and the protocol-level invariant that exposes it.
        To replay a trace, the agent overwrites the follower's log suffix unconditionally (red); this passes
        trace validation but lets a delayed append-entries message truncate a committed entry,
        which violates State Machine Safety.
        The correct repair (green) truncates only on conflicts.}
    \label{fig:repair-commit}
\end{figure}

\subsubsection{Repairing models.}
\label{sec:conformance:repair}
The key challenge of repairing models that fail trace validation 
    comes from the fact that AI agents tend to overfit the model 
    to match traces by weakening guards, 
    adding permissive transitions, or even hardcoding trace-specific state updates.
For example, a voting action that allows a node to vote for any candidate 
    can make election traces replay successfully, but it admits behaviors that violate the voting rules in the implementation.
In essence, trace validation checks that code-level actions can occur in the model, 
    but it does not explore whether the model also admits illegal actions.
(This was less a problem in the past as human experts would not repair models by overfitting.)
\specula therefore uses bidirectional validation to converge the model and the code 
    toward conformance.

Our insight is that model checking can be used to restrict the model correctness
    and prevent overfitted repairs.  
\specula model-checks the repaired model against protocol-level invariants (\S\ref{sec:inv}),
    to capture incorrect overfitted repairs.
Figure~\ref{fig:repair-commit} shows an example from Kudu-Raft,
    where model-checking the repaired model against an invariant on State Machine Safety
    exposes an overfitted repair.


Note that \specula does not assume AI-generated invariants and repaired models are always correct;
    Section~\ref{sec:evo-loop} describes how \specula handles those issues.


\subsection{Finding and reproducing buggy code behavior}
\label{sec:findbugs}

After achieving model-code conformance,
    \specula model-checks each scenario-based model against
    the protocol- and code-level invariants under the fault model.
An invariant violation may indicate a bug in the code.

\specula first runs TLC in breadth-first mode~\cite{yu99tlc},
which exhaustively covers every behavior up to a predefined depth.
Many bugs lie deeper than breadth-first search can reach within the time budget.
\specula therefore adds TLC's simulation mode~\cite{yu99tlc},
    which samples long random traces.

\subsubsection{Reproducing buggy code behavior.}
\specula attempts to reproduce buggy code behavior using the model trace
    and encapsulate the reproduction in a test.
Since the model trace describes the exact sequence of events that triggers the invariant violation,
    reproduction is not a search over any schedules
    but a controlled task of forcing one known schedule onto a code execution.

\specula instructs the agents to control (1) faults and (2) the order of events (controlled
    by adding sleep between events).
Specifically, the AI agent writes a test, observes the order of events,
    and compares it against the target sequence in the trace.
The procedure is tedious but mechanical.
\specula enforces a structured search by instructing agents to
    reproduce the behavior in four phases:
(1) interacting with the system via its client APIs,
(2) adding sleep between the API calls to control concurrency externally,
(3) crafting preconditions in terms of system state (e.g., a value left by a crash fault), and
(4) adding sleep in system code to control internal concurrency.
The agent stops at the phase that can reproduce the target behavior.
\specula forbids shortcuts that would craft the violation
    such as preloading illegal state,
    directly calling private functions,
    or changing system code logic.

For example, \specula can deterministically reproduce a deadlock
    in libgomp (GNU Offloading and Multi Processing Runtime Library).
Model checking finds the thread schedule that triggers the bug:
    an external thread completes a detached task only after every other thread
    has parked in the barrier's wait loop,
    at which point the wake-up fails to mark the barrier's task as
    pending and no thread can finish the barrier.
\specula inserts a sleep in the external thread with precise timing,
    turning a race into a deterministic deadlock.
(This bug was introduced in 2021 and was only found by \specula.)

In the case where \specula failed to reproduce the buggy behavior after the
    four phases of attempts,
    \specula instructs the agents to understand the reason based on
    the code-level traces produced during the reproduction.
If the agent thinks that the model trace cannot be reproduced
    at the code level (e.g., certain branch conditions are missed in the model),
    \specula restarts model repair (\S\ref{sec:conformance:repair}).
If the agent thinks the reason is the difficulty of deterministic replay
    (e.g., it requires external control), \specula preserves the
    analysis and evidence for the user.
Empirically, \specula reliably reproduced \draftnum{98}\% of the violations
    it judged to be real bugs.

\subsubsection{Comprehending buggy behaviors.}
\label{sec:consequence}
\specula observes the system-level impact of buggy behaviors
    to understand their consequences.
\specula first checks whether the reproduced execution leads to an observable
    consequence such as data loss, a crash, a deadlock, or a wrong output.
If none occurs, \specula does not discard the violation;
it instructs the agents to read issue discussions, design documents,
    tests, and comments, and interprets the violation in that context.
In our experience, \specula's prediction of the consequence is reasonable,
    interpretable, and evident.

\subsection{Self-evolving loops}
\label{sec:evo-loop}

We have not systematically discussed how to address mistakes 
    or errors introduced by AI agents---every component in Sections~\ref{sec:inv}--\ref{sec:findbugs}
    is powered by AI agents and thus is subject to hallucinations~\cite{ji2023hallucination,xu2024inevitable}
    and knowledge gaps (e.g., due to limited context windows)~\cite{liu2024lostmiddle,hsieh2024ruler}.

\specula does not assume the inherent correctness of any outputs of AI agents.
Instead, it carefully devises self-evolving loops that enable AI agents to correct erroneous
    invariants, formal models, and code instrumentation.
As a key design principle, for each loop iteration, \specula enforces 
    AI agents to collect new information and reason about their outputs with
    more evidence, thus gradually enhancing the agents' understanding of the target system.
In this way, the agents are improving and unlikely to make similar mistakes.

\subsubsection{Loop structure.}

\begin{figure}[t]
    \centering
    \includegraphics[width=\columnwidth]{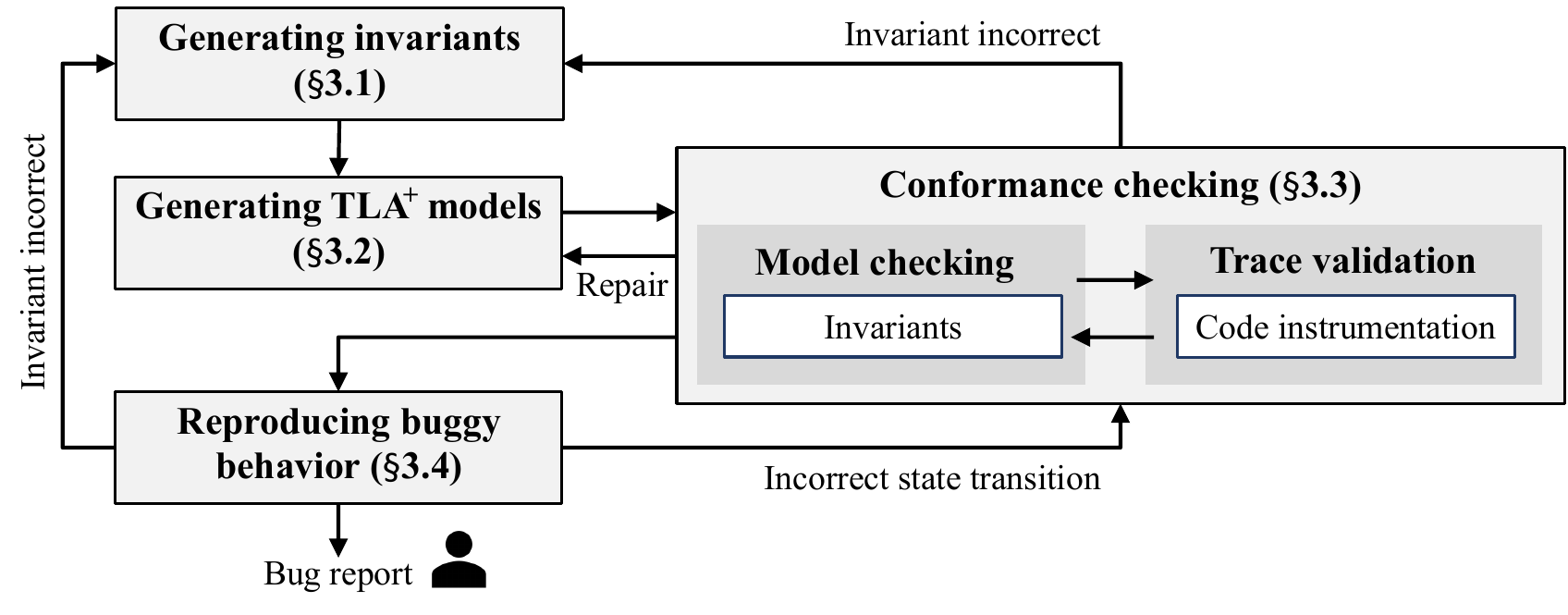}
     \vspace{-17.5pt}
    \caption{\specula's self-evolving loops}
    \label{fig:evo-loops}
\end{figure}

Figure~\ref{fig:evo-loops} shows the self-evolving loops across 
    the components in \specula.
We mainly discuss two interdependent loops that
    involve model-code conformance (\S\ref{sec:conformance})
    and buggy behavior reproduction (\S\ref{sec:findbugs}).

\para{Conformance loops.}
To achieve model-code conformance, the AI agents continuously 
    evolve the reference and scenario-based models 
    through a loop of trace validation and model checking.
The former ensures that code-level behaviors are permitted by the model,
    while the latter prevents agents from repairing models by overfitting
    code-level traces (\S\ref{sec:conformance}).
With the right invariants, the agents would evolve until the generated \tla{} models
    permit all code-level traces and satisfy protocol-level invariants
    (or it finds a code bug).

In a tricky situation, when a repaired model fails a protocol-level invariant,
    there can be a combination of three cases:
    (1) the model is still incorrect,
    (2) there is a code bug that fails the invariant, and
    (3) the invariant is incorrect.
\specula instructs the agent to reason about the three cases
    and decide the corresponding action:
    (1) continuing to repair the model,
    (2) reproducing the bug (see \S\ref{sec:findbugs}),
    and/or (3) correcting the invariant, 
        and then redoing model generation and conformance checking (\S\ref{sec:inv}--\S\ref{sec:conformance}).
In our experience, agents typically judge (3) correctly, when 
    enforced by \specula to provide concrete evidence
    from code, comments, and other artifacts as justifications.
In fact, if agents make mistakes on invariants, 
    the process will not
    converge but continue to iterate.
On the other hand, agents often have difficulty differentiating (1) and (2),
    as both surface a model state that violates invariants---the code alone does not directly show whether the state 
    can be reached at runtime.
If the agent mistakenly treats (2) as (1), it will tighten the model to exclude
    the state, which will later be rejected by trace validation.
If the agent mistakenly treats (1) as (2), the mistake will be surfaced during 
    bug reproduction.

\para{Reproduction loops.}
\specula reproduces buggy code behavior event by event (\S\ref{sec:findbugs}).
If the agents cannot reproduce events,
    it restarts the conformance loops with the diverging states as the feedback
    (Figure~\ref{fig:evo-loops}): it is possible that conformance checking missed
    rare cases due to its incompleteness.
If reproduction succeeds but no observable consequence occurs (\S\ref{sec:consequence}),
    it indicates three cases: 
    (1) the invariant is overly strong,
    (2) the consequence is masked by the system's built-in mitigation/recovery modules,
    and (3) the agent fails to reproduce the bug.
\specula instructs the agents to reason about the three cases 
    and choose to (1) improve the invariant, 
        which may trigger a redo of model generation and conformance checking (\S\ref{sec:inv}--\S\ref{sec:conformance}),
    and (2) provide evidence and report to users (masked violations still indicate bugs),
    and (3) report to users and let them judge.

\subsubsection{Correctness.} 
    The evolving loops are safe---they should not incorrectly reduce
    \specula's bug-finding capabilities.
First, the loop between trace validation and model checking
    only gradually tightens model-code conformance
    without relaxing invariants.
Second, for the conformance loop
    that revises protocol-level invariants,
    the updated invariants are expected to become more precise, 
    with a detailed understanding of why the original protocol-level invariants
    are incorrect.
Third, for the reproduction loop that revises invariants,
    the updated invariants are expected 
    to reflect code-level correctness properties more closely.
Note that revising invariants requires AI agents to provide strong evidence
    that the original invariants were incorrect or unnecessary;
    it is not used as a way to work around violations.


With the assumption that 
    agents improve over the iterations, \specula offers convergence---the invariants
    eventually converge and the agents eventually evolve the \tla{} model 
    that conforms to the system code.
In practice, we run \specula under a budget in time or cost.
In our evaluation, end-to-end runs take 1.43 to 9.86 hours
    across the target systems.

\section{Implementation}
\label{sec:implementation}

\specula is built on coding agents such as Claude Code, Codex,
    and Copilot CLI. 
It includes about 5.5K lines of Markdown Agent Skills and 
    24.8K lines of code across Python, UNIX shell, and Java,
    for supporting tools and orchestration.

Each component
     (see Figure~\ref{fig:evo-loops}) is implemented as a multi-agent workflow:
    a primary coding agent equipped with specific skills and tools, who may 
    spawn more agents for scoped subtasks. 
The components are orchestrated as producers and consumers---a component
    can consume outputs (in file format) produced by other components.
These files can be used as checkpoints for retrying a component 
    or forking multiple instances of a component (e.g., for processing 
    multiple specification variants).
Each component runs coding agents inside a shared workspace 
    that contains the target system artifact
    and component-specific workspaces.

\para{Tools.} Besides standard CLI tools like UNIX shells,
    \specula equips its agents with \tla{} tools:
    (1) the SANY parser~\cite{tlaplus}, (2) the
    TLC model checker~\cite{yu99tlc}, and a static analyzer we wrote
    for checking if each action
    correctly specifies an atomic transition over all declared variables (an important
    feature missed by SANY and TLC~\cite{tlaplus_unchanged_issue}).
Together, these tools let agents catch and correct common mistakes
    such as missing \textsf{UNCHANGED} clauses, duplicate assignments,
    inconsistent action structures, and malformed configuration files.

For code instrumentation, \specula offers the agent a trace library
    we wrote.
The trace library exposes per-language emit functions for agents 
    to insert at each instrumentation point to record a trace event.
The agent uses the trace library with the system's build and test tools
    to insert trace events in the code.
For distributed systems, \specula uses a mutex-based recording mechanism,
    following prior work~\cite{ouyang2025remix,tang2024sandtable,cirstea2024trace};
for concurrent systems, \specula uses a
    timebox-based recording mechanism~\cite{hackett2026omnilink} that captures 
    per-thread operation intervals and lets validation search for an
    ordering consistent with the intervals, minimizing the impact
    of instrumentation.

For trace validation, \specula offers the agent a trace inspector
    and a debugger.
The inspector provides APIs for agents to query events, variables, and time windows
    and return matching states with optional in-situ evaluation of 
    specification expressions on that state.
A trace may contain thousands of lines; loading the full trace 
    into the agent's context is inefficient.
Hence, we wrote a debugger~\cite{dap} that supports breakpoints, hit counts, expression evaluation, and
    variable inspection, letting the agent pause validation at any point and
    inspect any specification expression or runtime state.
This is useful, because raw trace validation reports only the depth at
    which TLC stops, not the failure state; also, the true
    divergence may originate in an auxiliary variable that drifted many
    actions before the reported point. 
The debugger lets the agent walk back to the specific transition where the
    divergence began and trace how it propagated through subsequent actions until the violation.

\section{Evaluation}
\label{sec:eval}

Our evaluation aims to answer the following questions:
\begin{enumerate}[leftmargin=*]
    \item Is \specula effective in finding important, deep bugs?
    \item How do \specula's capabilities of specification and bug-finding compare to standard agentic approaches?
    \item What is the cost of using \specula to check a system?
    \item How effective are \specula's self-evolving loops (\S\ref{sec:evo-loop}) in correcting mistakes?
    \item How sensitive is \specula's effectiveness to the capabilities of the coding agent it uses?
\end{enumerate}
We apply \specula to \numcases{} system projects, including \numdistributed{} distributed systems
    and \numconcurrency{} concurrent systems.
The evaluated projects include widely used open-source projects such as
    MongoDB and GCC libgomp, as well as artifacts from
    research papers like Autobahn~\cite{giridharan2024autobahn}.
A full list of evaluated systems can be found in Table~\ref{tab:eval-bugs}.
These systems span seven different languages (including C, C\#, C++,
    Erlang, Go, Java, and Rust) with various sizes (\locmin{} to \locmax{} LoC).

For all the experiments, we ran \specula with Claude Code v2.1.97 (Claude Opus-4.8, 1M context window, 
    with \texttt{max} reasoning)
    on an Azure VM with a 96-core AMD EPYC 9V74 CPU and 384~GB of RAM.
Each run executes in an isolated git workspace that symlinks to the target system project.

\subsection{Bugs found}
\label{sec:eval-bugs}

We treat bug finding as one measure of \specula's utility.
We have applied \specula to \numcases{} system projects in the course 
    of its development (many were by requests).
In total, \specula found \numbugs{} bugs in the \numcases{} systems:
    \numnewbugs{} were {\it new} bugs 
    and \numknownbugs{} were known bugs (but not fixed).
We reported \numreported{} bugs;
    so far, \numconfirmed{} have been confirmed and \numfixed{} have been fixed. 
Table~\ref{tab:eval-bugs} shows the number of bugs found by \specula
    per system.
Note that only the highlighted results reflect the latest \specula
    (our v1.0 release); the other results are from early versions 
    of \specula; we expect the latest \specula to find more bugs
    (we are limited by our budget to rerun everything).

As \specula is instructed 
    to target core logic of concurrent, distributed systems, 
    which requires formal methods and model checking,
    the bugs found by \specula are deep and have severe consequences,
    including:
\begin{itemize}[leftmargin=*]
    \item {\bf Deadlocks and system hangs}, e.g., GCC's libgomp can run into deadlocks 
        due to a missing flag on a barrier wake-up path~(\S\ref{sec:libgomp-case-study});
    \item {\bf Data loss or corruption}, e.g., MongoDB's sharding
      module marks the wrong migration task as ready, leading to incorrect data deletion;
    \item {\bf Crashes and component failures}, e.g., ra, the Raft library of RabbitMQ's quorum queues,
        crashes a follower process when a new leader overwrites an uncommitted membership change on a regular log entry;
    \item {\bf Loss of availability}, e.g., in HashiCorp's Raft library,
      which underlies Consul and Vault, a leader whose disk has stalled keeps
      sending heartbeats that suppress elections, so the cluster can neither
      commit nor fail over.
\end{itemize}


\begin{table}[t]
\centering
\footnotesize
\caption{The numbers of bugs found by \specula in each evaluated system.
    ``New'' refers to new bugs and the numbers in ``()'' refer to bugs
    that were confirmed by the developers. The shaded rows are 
    results from the latest \specula version.}
\vspace{-8pt}
\label{tab:eval-bugs}
\setlength{\tabcolsep}{5pt}
\begin{tabular}{@{}l l l r r@{}}
\toprule
\textbf{System} & \textbf{Type} & \textbf{Lang.} & \textbf{New} & \textbf{Total} \\
\midrule
\rowcolor{gray!15}\href{https://github.com/aptos-labs/aptos-core}{Aptos Quorum Store} & BFT & Rust & 2 (0) & 2 \\
\rowcolor{gray!15}\href{https://github.com/aptos-labs/aptos-core}{AptosBFT} & BFT & Rust & 1 (0) & 1 \\
\href{https://github.com/vorner/arc-swap}{arc-swap} & Concurrency & Rust & 2 (1) & 7 \\
\href{https://github.com/async-raft/async-raft}{async-raft} & Raft & Rust & 8 (0) & 8 \\
\rowcolor{gray!15}\href{https://github.com/neilgiri/autobahn-artifact}{Autobahn} & BFT & Rust & 16 (15) & 16 \\
\rowcolor{gray!15}\href{https://github.com/babylonlabs-io/babylon}{Babylon} & BFT & Go & 4 (0) & 6 \\
\href{https://github.com/hyperledger/besu}{Besu QBFT} & BFT & Java & 2 (1) & 3 \\
\href{https://github.com/baidu/braft}{braft} & Raft & C++ & 4 (0) & 5 \\
\rowcolor{gray!15}\href{https://github.com/cometbft/cometbft}{CometBFT} & BFT & Go & 10 (0) & 15 \\
\href{https://github.com/crossbeam-rs/crossbeam}{crossbeam-deque} & Concurrency & Rust & 2 (1) & 2 \\
\href{https://github.com/crossbeam-rs/crossbeam}{crossbeam-epoch} & Concurrency & Rust & 0 (0) & 2 \\
\href{https://github.com/crossbeam-rs/crossbeam}{crossbeam-skiplist} & Concurrency & Rust & 1 (0) & 2 \\
\href{https://github.com/dotnet/dotNext}{dotNext} & Raft & C\# & 4 (2) & 4 \\
\href{https://github.com/DPDK/dpdk}{dpdk-ring} & Concurrency & C & 1 (0) & 1 \\
\href{https://github.com/lni/dragonboat}{Dragonboat} & Raft & Go & 1 (0) & 1 \\
\href{https://github.com/eliben/raft}{eliben-raft} & Raft & Go & 4 (3) & 4 \\
\href{https://github.com/imdea-software/swiftpaxos}{Epaxos} & Paxos & Go & 3 (0) & 3 \\
\href{https://github.com/etcd-io/raft}{etcd/raft} & Raft & Go & 1 (0) & 2 \\
\rowcolor{gray!15}\href{https://github.com/gcc-mirror/gcc}{GCC libgomp} & OpenMP & C & 3 (2) & 3 \\
\rowcolor{gray!15}\href{https://github.com/algorand/go-algorand}{go-algorand} & BFT & Go & 2 (0) & 2 \\
\href{https://github.com/goraft/raft}{goraft} & Raft & Go & 5 (0) & 5 \\
\href{https://github.com/hashicorp/raft}{HashiCorp Raft} & Raft & Go & 2 (2) & 3 \\
\rowcolor{gray!15}\href{https://github.com/EspressoSystems/HotShot}{HotShot} & BFT & Rust & 5 (0) & 5 \\
\href{https://github.com/fereidani/kanal}{kanal} & Concurrency & Rust & 3 (0) & 3 \\
\href{https://github.com/jonhoo/left-right}{left-right} & Concurrency & Rust & 2 (0) & 3 \\
\rowcolor{gray!15}\href{https://github.com/DMTF/libspdm}{libspdm} & SPDM & C & 25 (16) & 26 \\
\href{https://github.com/llvm/llvm-project}{LLVM libomp} & OpenMP & C++ & 1 (0) & 1 \\
\rowcolor{gray!15}\href{https://github.com/mongodb/mongo}{MongoDB} & Database & C++ & 15 (0) & 15 \\
\href{https://github.com/imdea-software/swiftpaxos}{N2Paxos} & Paxos & Go & 3 (0) & 3 \\
\href{https://github.com/vesoft-inc/nebula}{nebula (Raft)} & Database & C++ & 2 (0) & 2 \\
\href{https://github.com/eBay/NuRaft}{NuRaft} & Raft & C++ & 3 (0) & 3 \\
\href{https://github.com/ibraheemdev/papaya}{papaya} & Concurrency & Rust & 4 (1) & 4 \\
\href{https://github.com/rabbitmq/ra}{rabbitmq/ra} & Raft & Erlang & 7 (7) & 7 \\
\href{https://github.com/wenweihu86/raft-java}{raft-java} & Raft & Java & 2 (0) & 5 \\
\href{https://github.com/RedisLabs/redisraft}{RedisRaft} & Raft & C & 2 (0) & 2 \\
\href{https://github.com/rethinkdb/rethinkdb}{RethinkDB (Raft)} & Database & C++ & 1 (1) & 2 \\
\href{https://github.com/wvwwvwwv/scalable-concurrent-containers}{scc} & Concurrency & Rust & 0 (0) & 1 \\
\href{https://github.com/scylladb/scylladb}{ScyllaDB (Raft)} & Database & C++ & 1 (1) & 1 \\
\rowcolor{gray!15}\href{https://github.com/sofastack/sofa-jraft}{sofa-jraft} & Raft & Java & 10 (3) & 10 \\
\rowcolor{gray!15}\href{https://github.com/anza-xyz/agave}{Solana} & BFT & Rust & 0 (0) & 8 \\
\rowcolor{gray!15}\href{https://github.com/sonic-net}{SONiC} & Network OS & C/C++/Rust & 19 (1) & 25 \\
\href{https://github.com/ccc-spdm-tools/spdm-rs}{spdm-rs} & SPDM & Rust & 10 (8) & 10 \\
\href{https://github.com/paritytech/substrate}{Substrate} & BFT & Rust & 0 (0) & 1 \\
\rowcolor{gray!15}\href{https://github.com/MystenLabs/sui}{Sui} & BFT & Rust & 2 (0) & 2 \\
\href{https://github.com/imdea-software/swiftpaxos}{SwiftPaxos} & Paxos & Go & 2 (2) & 2 \\
\href{https://github.com/tikv/raft-rs}{tikv/raft-rs} & Raft & Rust & 3 (1) & 4 \\
\href{https://github.com/tokio-rs/tokio}{tokio-broadcast} & Concurrency & Rust & 2 (0) & 2 \\
\href{https://github.com/willemt/raft}{willemt/raft} & Raft & C & 5 (0) & 5 \\
\midrule
\textbf{Total} & & & \textbf{207 (68)} & \textbf{249} \\
\bottomrule
\end{tabular}
\end{table}

\begin{figure}[t]
    \centering
    \includegraphics[width=\linewidth]{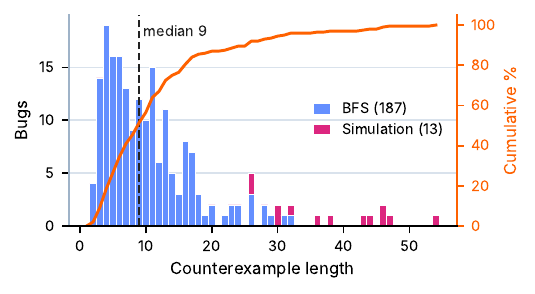}
    \vspace{-25pt}
    \caption{Counterexample lengths of the \mcbugs{} bugs found via
      model checking. BFS returns the shortest counterexample;
      Simulation does not. The CDF curve covers all \mcbugs{} bugs.}
    \label{fig:bfs-depths}
\end{figure}

Among the \numbugs{} bugs \specula found, \mcbugs{} (80.3\%) 
    surfaced through model checking;
    the rest were found during code comprehension and modeling only.
Of the \mcbugs{} bugs surfaced by model checking, \bfsbugs{} were found by
    breadth-first search (BFS), which returns the shortest counterexample.
Figure~\ref{fig:bfs-depths} plots the distribution of these counterexample lengths
    with a median of \cexmedian{} steps and a p90 of 18 steps.
The other \simbugs{} were found by random simulation whose traces are not
    the shortest.
BFS sufficed for \bfsbugs{} (93.5\%) of the \mcbugs{} violations, showing that \specula's
    scenario-based decomposition keeps each model's state space tractable enough for model
    checking to reach the bug within a bounded time budget.

Most (\pctsafety) invariants \specula generates across the \numcases{} systems are safety properties.
Only \pctliveness{} are liveness properties.
\specula mostly checks liveness through TLC's deadlock detection, which
    works in both BFS and random simulation.
For a small number of models whose state space can be exhausted, 
    \specula checks the temporal formulas directly, 
    reporting a liveness violation when the model's fairness
    assumptions still admit an execution that keeps taking steps but never makes progress.
Table~\ref{tab:invariants} shows the percentages of sources \specula derives invariants from.
Among these invariants, \pctprotocol{} are at the protocol level and
    \pctcodelevel{} are at the code level (\S\ref{sec:inv}).

\subsubsection{False positives.}
For the \draftnum{14} systems checked with the latest \specula (shaded in
    Table~\ref{tab:eval-bugs}), \specula reported \draftnum{136} bugs.
For \draftnum{134} of them, \specula reproduced the violations at the code level
    and encoded them in a test (\S\ref{sec:findbugs}).
For the remaining two bugs, \specula also reproduced both of them,
    but did not observe severe consequences; so, \specula considered them 
    as bugs that were masked by the system.


\begin{table}[t]
\centering
\small
\caption{Distribution of sources of evidence \specula cited when generating 
  invariants across evaluated projects. One invariant can have multiple sources.}
\label{tab:invariants}
\vspace{-8pt}
\begin{tabular}{l r}
\toprule
\textbf{Source of evidence} & \textbf{Invariants} \\
\midrule
Implementation code and comments                 & \srccode \\
Issue trackers, pull requests, and advisories    & \srcissues \\
Prior fixes, commit history, and tests           & \srcfixes \\
Documentation and reference literature           & \srcdocs \\
\bottomrule
\end{tabular}
\end{table}

\subsubsection{Case studies}
\label{sec:libgomp-case-study}
We present two case studies, where we use \specula to check 
    GCC libgomp and SONiC, which represent real-world
    concurrent and distributed systems.

\para{GCC libgomp.}
The GNU Offloading and Multi-Processing library (aka libgomp)
    is the runtime library of GCC for 
    shared-memory parallel programming.
Its low-level concurrency is hard to reason about and the developers 
    asked us to help model-check it using \specula.
We ran \specula on an unmerged patch that adds a faster
    barrier, 
    directing the agent to focus on it.
\specula surfaced two bugs (Figure~\ref{fig:libgomp}): one introduced by the patch and one
    latent in an existing barrier.
The run took about six hours of compute and 1.5 hours of human review to confirm the bugs.

\begin{figure}[t]
\centering
\begin{subfigure}{\columnwidth}
\begin{minted}{c}
gomp_increment_gen(unsigned gen, int incr) {
  gens = gen & BAR_BOTH_GENS_MASK;  // flags wiped
  ...
  case BAR_HOLDING_SECONDARIES:
    return gens | BAR_HOLDING_SECONDARIES
      | (gen & BAR_CANCELLED);  // keep cancel
}
\end{minted}
\par\vspace{2pt}
{\footnotesize\raggedright
\textcolor{red}{\ding{55}}~{\bf Safety invariant:} A \textsf{BAR\_CANCELLED}
  flag, which was set, must be carried by the barrier's advance to the next round.}\par
\caption{Cancellation flag dropped when the barrier finishes.}
\label{fig:libgomp-cancel}
\end{subfigure}

\vspace{6pt}

\begin{subfigure}{\columnwidth}
\begin{minted}{c}
omp_fulfill_event(omp_event_handle_t event) {
  if (new_tasks > 0) {  // path 1
    gomp_team_barrier_set_task_pending(...);
    do_wake = ...; }
  if (!do_wake && waiting_for_tasks(...)) {
    gomp_team_barrier_set_task_pending(...);
    do_wake = 1; }
}
\end{minted}
\par\vspace{2pt}
{\footnotesize\raggedright
\textcolor{red}{\ding{55}}~{\bf Liveness invariant:} Any wake-carrying
  task must set the pending flag; otherwise the woken thread sleeps and causes deadlocks.\par}
\caption{A wake path omits \textsf{set\_task\_pending}.}
\label{fig:libgomp-wake}
\end{subfigure}

\caption{Two bugs \specula found in libgomp and the code-level invariants they violate.}
\label{fig:libgomp}
\end{figure}

The first bug (Figure~\ref{fig:libgomp-cancel}) resides in the
    fast barrier implementation.
libgomp encodes the barrier's round counter and its status flags in an
    integer; one of those flags is for pending cancellation (\textsf{BAR\_CANCELLED}).
    When the barrier advances to the next round, a helper recomputes this integer from the
    counter bits alone, wiping every flag, then restores only the flag that the increment sets.
    However, it never restores \textsf{BAR\_CANCELLED}, so a cancellation in
    flight at that moment reaches only some of the threads. Those threads stop, while the
    rest see the barrier complete normally and keep running. The result is a split-brain scenario,
    and the patch developer confirmed and fixed this bug.

The second (Figure~\ref{fig:libgomp-wake}) is a deadlock that had been latent in
    the codebase for at least five years. 
For the two paths that wake a barrier thread, one
    omits the call that marks tasks as pending,
    so a woken thread can return to sleep with work still queued.
The same developer confirmed it. 

Both bugs surface only under specific thread interleavings.
The deadlock needs a parked thread to be woken because work is still
    outstanding, rather than through the common path taken when a new task is
    created.
The cancellation bug needs a cancellation in flight at the moment the leader
    thread advances the barrier and pauses the other threads.
Such interleavings are hard for ordinary tests to produce, so the bugs went
    undiscovered; model checking enumerates all interleavings and captures
    the triggering path.%

\para{SONiC.}
SONiC is a network operating system used primarily for switches in datacenter
    networks.
It has multiple modules, from programming the
    switch's forwarding hardware to running peering protocols to recovering state
    after reboots.
We ran \specula on five modules, written in C++, C, and Rust, with
    no module-specific tuning.
\specula produced \tla{} specifications for each and the model checking
    surfaced at least one bug in each module (see Table~\ref{tab:sonic-bugs}).

Figure~\ref{fig:sonic} shows two bugs.
In Figure~\ref{fig:sonic-iccpd}, iccpd keeps two switches in sync: a handshake
    first advances a state machine into the operational \textsf{EXCHANGE} state.
    The handler that sends a peer its sync data ends by advancing this state machine, 
    which is correct during the handshake but also runs during a subsequent re-sync.
    Taken in \textsf{EXCHANGE}, that extra advance overshoots into an
    \textsf{ERROR} state with no recovery, and the switches stop coordinating
    until the connection resets.

In Figure~\ref{fig:sonic-dash-ha}, dash-ha notifies every dependent actor when initiating an HA set, 
    but incorrectly notifies no actor when tearing the HA down.
    As a consequence, the actors rely on an HA set that no longer exists.
    Neither bug is reachable by the project's tests (the former needs a precise message
    ordering, the latter needs a test that exercises the teardown path, which does not exist).
    Model checking finds both and \specula reproduces them at the code level.

\begin{table}[t]
\centering
\footnotesize
\setlength{\tabcolsep}{4pt}
\caption{Examples of bugs \specula found in SONiC modules}
\label{tab:sonic-bugs}
\vspace{-8pt}
\begin{tabular}{@{}p{1.4cm} p{3.7cm} p{3.0cm}@{}}
\toprule
\textbf{Modules} & \textbf{Bug} & \textbf{Consequence} \\
\midrule
dash-ha {\scriptsize(Rust)}   & Teardown of a group misses notifications to components using it & The components keep acting on deleted state \\
\midrule
fdb {\scriptsize(C++)}        & Removing a lookup entry before hardware invalidation & Inconsistency between software and hardware \\
\midrule
iccpd {\scriptsize(C)}        & An incorrect sync drives the protocol into an error state & The coordination of two switches fails \\
\midrule
linkmgrd {\scriptsize(C++)}   & A startup case is missing from the failover logic & The link is left stuck with no recovery \\
\midrule
warmreboot {\scriptsize(C++)} & Reporting recovery complete before a dependency is restored & Forwards traffic using state that is not yet rebuilt \\
\bottomrule
\end{tabular}
\end{table}

\begin{figure}[t]
\centering
\begin{subfigure}{\columnwidth}
\begin{minted}{c}
mlacp_sync_send_all_info_handler(csm) {
  ...  // first send all sync data to the peer
  if (MLACP(csm).current_state < EXCHANGE)
    MLACP(csm).current_state++;  // stay in FSM
}
\end{minted}
\par\vspace{2pt}
{\footnotesize\raggedright
\textcolor{red}{\ding{55}}~{\bf Safety invariant:} once the two peers
  finish the handshake (state \textsf{EXCHANGE}), a later sync request
  must not change the protocol state.\par}
\caption{In iccpd, a re-sync after the handshake turns the state machine
  into an unrecoverable \textsf{ERROR} state.}
\label{fig:sonic-iccpd}
\end{subfigure}

\vspace{6pt}

\begin{subfigure}{\columnwidth}
\begin{minted}{rust}
// setup path notifies every scope actor:
notify_scope(HaSetState { up: true });
// teardown must send the matching down:
notify_scope(HaSetState { up: false });
\end{minted}
\par\vspace{2pt}
{\footnotesize\raggedright
\textcolor{red}{\ding{55}}~{\bf Safety invariant:} tearing down an HA set
  must notify every actor that its setup did, so none keeps a stale \textsf{up}
  state.\par}
\caption{In dash-ha, teardown misses sending notifications to actors 
    that were previously notified during the HA setup.}
\label{fig:sonic-dash-ha}
\end{subfigure}

\caption{Two bugs \specula found in SONiC, which violate protocol-level
  (iccpd) and code-level (dash-ha) invariants.}
\label{fig:sonic}
\end{figure}

\subsection{Comparison with other agentic approaches}
\label{sec:eval-baselines}

We compare \specula against two agentic approaches as our baseline (all three 
    approaches use the same prompts):
\begin{itemize}[leftmargin=*]
    \item \textbf{Agent-Raw.} Claude Code without additional Skills or MCPs. 
        Agent-Raw represents the raw LLM's ability without using domain-specific tools.
    \item \textbf{Agent-\tla{}.} Claude Code equipped with the \tla{} Skills and MCPs~\cite{tlaplus_agentskills}.
        Agent-\tla{} represents official tools for syntax checking and model checking.
        Unlike \specula, Agent-\tla{} is not designed
        with techniques for model-code conformance or self-evolving loops
        to correct model or invariant errors. 
\end{itemize}
For this analysis,
    we select Autobahn, CometBFT, libspdm, MongoDB, and sofa-jraft---written in
    Rust, Go, C, C++, and Java respectively (see Table~\ref{tab:eval-bugs}).
We compare the quality of the generated \tla{} models,
    the bugs found, and the false positives of the three approaches.

\begin{table}[t]
\centering
\small
\caption{Specification quality measured by SysMoBench~\cite{cheng2026sysmobench}.}
\label{tab:sysmobench}
\vspace{-8pt}
\setlength{\tabcolsep}{2pt}
\begin{tabular}{l ccccc}
\toprule
\textbf{Agent} & \textbf{Syntax} & \textbf{Runtime} & \textbf{Conform.} & \textbf{Inv.} & \textbf{Overall} \\
\midrule
\specula{} (Opus-4.8)             & 100\% & 100\% & 100\% & 100\% & 100\% \\
Agent-Raw     & 100\% & 79\% & 70\% & 84\% & 81\% \\
Agent-\tla{}  & 100\% & 82\% & 75\% & 82\% & 82\% \\
\midrule
\specula{} (Sonnet-4.6)      & 100\% & 100\% & 90\% & 90\% & 93\% \\
\specula{} (Haiku-4.5)       & 95\% & 56\% & 51\% & 17\% & 47\% \\
\bottomrule
\end{tabular}
\end{table}

\subsubsection{Quality of the generated \tla{} model.}
\label{sec:model-quality}
We use SysMoBench~\cite{cheng2026sysmobench} to evaluate the quality of the
    reference model generated by \specula and the other two agentic approaches (see \S\ref{sec:base-model}).
SysMoBench requires human-verified invariants as inputs and thus cannot 
    evaluate invariants; we manually wrote the invariants
    for the five evaluated systems.
SysMoBench introduces four metrics of model quality: 
    (1) syntax correctness, (2) runtime correctness (if the \tla{} model could run without exceptions),
    (3) model-code conformance, and (4) invariants (if the generated model satisfies invariants),
    with an aggregated overall score.

Table~\ref{tab:sysmobench} shows the results.
\specula achieves perfect scores on SysMoBench,
    while Agent-Raw and Agent-\tla{} do not.
All three approaches write perfect \tla{} syntax.
However, Agent-Raw and Agent-\tla{} cannot ensure 
    runtime correctness or model-code conformance
    (Agent-\tla{} achieves higher scores by using tools).
Specifically, we find that the official \tla{} tools bring no real improvement overall:
what is lacking is not \tla{} knowledge but the runtime feedback that 
    helps the agents to repair and revise the models.

\subsubsection{Bugs found.}

\begin{figure}[t]
\centering
\includegraphics[width=0.576\columnwidth]{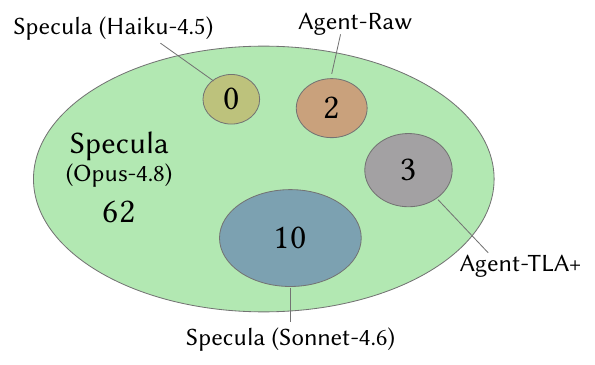}
\vspace{-12pt}
\caption{Comparative analysis of bug finding of different approaches.
    The default LLM is Claude Opus-4.8.}
\label{fig:rq2-bugs}
\end{figure}

As shown in Figure~\ref{fig:rq2-bugs}, \specula finds \gtbugs{} bugs, while 
    Agent-Raw finds \gtbasic{} and Agent-\tla{} finds \gttla{}.
The reason \specula found more bugs is threefold.
First, \specula builds more effective \tla{} models that describe important code-level behaviors
    related to the invariants, while the other approaches often abstract important behaviors out.
Second, \specula generates models for relevant
    scenarios, letting model checking explore the state space more effectively.
Third, \specula models more invariants, checking correctness properties the others do not.
The results show the importance of effective, targeted formal specifications.

\subsubsection{False positives.}
\label{sec:eval:comp:fp}
\specula reports no false positives; 
    Agent-Raw reports five and Agent-\tla{} reports two.
The false positives reported by Agent-Raw and Agent-\tla{} 
    all come from incorrect specifications when checking libspdm.
In one case, Agent-Raw generates an invariant stronger than the implementation guarantees, assuming
    a certificate is verified where the protocol makes no such promise, 
    so model checking reports a violation which is not a bug.
In another case, an Agent-Raw-generated model
    admits a behavior the code rules out, such as a message arriving
    after a client has unsubscribed.
libspdm's behaviors hinge on small, scattered details
    which are more fine-grained than in the other projects.
Modeling without grounding against the code, Agent-Raw and Agent-\tla{} drop these
    details and report violations that those details would rule out in the
    implementation.

\subsection{Cost}
\label{sec:eval-resources}

For the \numcases{} systems, \specula takes \runhoursmin--\runhoursmax{} hours 
    to check each system end-to-end, with a median of \runhoursmedian{} hours;
    these end-to-end checks cost \runcostmin--\runcostmax{} US dollars of token consumption,
    with a median of \runcostmedian{} US dollars.
We consider the monetary cost to be reasonable, considering the months of
    effort human experts need to invest in writing specifications by hand~\cite{howard2025ccf,davis2020extreme_modelling,cauli2026},
    let alone model checking, trace validation, and bug reproduction.
Reducing token cost is an active area of future work. For example, many
    subtasks do not need the strongest LLM and could be routed to cheaper LLMs.

\begin{table}[t]
\centering
\small
\setlength{\tabcolsep}{4pt}
\caption{Cost (in USD) to check a system project on average and 
    the corresponding running time (in minutes) for \specula,
    compared with Agent-Raw and Agent-\tla{}.}
\label{tab:rq3-cost-time}
\vspace{-8pt}
\begin{tabular}{l rr rr rr}
\toprule
 & \multicolumn{2}{c}{Agent-Raw} & \multicolumn{2}{c}{Agent-\tla{}}
   & \multicolumn{2}{c}{\specula} \\
\cmidrule(lr){2-3}\cmidrule(lr){4-5}\cmidrule(lr){6-7}
\textbf{System} & Time & Cost & Time & Cost & Time & Cost \\
\midrule
Autobahn   &  27 &  \$5.99 &  69 & \$16.46 & 110 &  \$28.96 \\
CometBFT   &  40 &  \$6.27 &  13 &  \$2.58 & 402 & \$167.83 \\
libspdm    &  14 &  \$3.81 &  16 &  \$5.08 & 151 &  \$40.75 \\
MongoDB    &  15 &  \$3.82 &  15 &  \$4.26 & 157 &  \$47.54 \\
sofa-jraft &  11 &  \$2.19 &  24 &  \$4.91 & 199 &  \$81.06 \\
\bottomrule
\end{tabular}
\end{table}

Table~\ref{tab:rq3-cost-time} shows the token cost and running time 
    for the five systems studied in \S\ref{sec:eval-baselines},
    compared with Agent-Raw and Agent-\tla{}.
Across the 19 modules of the five systems, \specula costs \costopus{}
    and takes 2.8 hours on average.
This is about 4.8--37$\times$ more expensive than Agent-Raw
    and 1.8--65$\times$ than Agent-\tla{}.
Most of the cost goes toward \tla{} model generation and code instrumentation.



We find that the token cost is more strongly correlated with
    the complexity of the target system than its size.
The cost of checking modules is not evenly distributed---a few complex modules 
    are significantly more expensive to check.
The two longest runs are CometBFT and libspdm's cert auth, which drive far larger searches
    than any other module.
For example, in CometBFT's PBTS module,
    the coding agent took 670 agent steps for checking and repairing the \tla{} model,
    more than 4$\times$ the other modules, and TLC generated 1.9 billion states
    while visiting 393 million distinct ones.

\subsection{Effectiveness of self-evolving loops}
\label{sec:eval-loops}

Agents routinely make mistakes.
For all the \numcases{} evaluated systems,
    \specula has to repair the \tla{} model it generates (\S\ref{sec:conformance:repair}).
For \numinvrevised{} systems, invariants were revised;
for \numinstrfixed{} systems, code instrumentation was corrected. 
Therefore, the self-evolving loops are essential.


The conformance loop caught and repaired three kinds of agent mistakes on
    invariants (\pcterrinv{}), code instrumentation (\pcterrinstr{}),
    and model (\pcterrmodel{}) (\S\ref{sec:evo-loop}).
Of the violations that entered reproduction, \pctconfrepro{} reproduced as real bugs,
    and \pctconfdischarged{} were discharged with evidence drawn from the system itself
    and fed back to repair the model or its invariants.
The agents reported the remaining cases to users; only \pctconfnorepro{} are considered bugs
    that \specula cannot reproduce.

All our recorded \specula runs converged.
Code instrumentation errors were repaired within three rounds, and \pctonefix{} of
    invariant and model errors were fixed in one iteration; none in more than four.
The one case that took four revisions came from SONiC's link manager, where 
    \specula generated an incorrect invariant that a link's two gateways are never both on standby.
Differentiating a bug from a legitimate transient is hard: in each round, model
    checking surfaced a different valid case, such as a failover handoff or a degraded
    mode where neither gateway can take over, and after four counterexamples the agent
    arrived at a correct invariant.

\subsection{Sensitivity to coding agents}
\label{sec:eval-models}

We measure \specula's sensitivity to the underlying coding agents 
    by applying \specula to the same five system projects in \S\ref{sec:eval-baselines},
    with two different LLMs, Claude Sonnet-4.6 and Haiku-4.5 (which have 
    weaker capabilities than Opus-4.8).

\para{Quality of \tla{} models.}
We use the same methodology to evaluate the quality of generated \tla{} models 
    as in \S\ref{sec:model-quality}.
Table~\ref{tab:sysmobench} shows the results.
\specula with Sonnet-4.6 achieved syntax and runtime correctness and its 
    overall quality is higher than those generated by
    Agent-Raw and Agent-\tla{} with Opus-4.8.
However, with Sonnet-4.6, \specula was not able to achieve 
    complete conformance and invariant correctness
    (90\% each): its model replays fewer of the implementation's traces, and
    some of the invariants fail on it even though the invariants it wrote hold.
With Haiku-4.5, \specula fails earlier.
\specula{} asks more of the LLM than plain specification generation, and with Haiku-4.5
    it already struggles to produce a model that runs: MongoDB's
    specification fails syntax correctness, and on average TLC can execute
    only about half of the actions in its models.

\para{Bugs found.}
With Sonnet-4.6, \specula finds \gtsonnet{} of the \gtbugs{} bugs it finds with Opus-4.8
    (Figure~\ref{fig:rq2-bugs}).
With Haiku-4.5, it finds none.
With Sonnet-4.6, \specula found bugs in Autobahn, CometBFT, and libspdm (none in
    MongoDB or sofa-jraft).
With Sonnet-4.6, \specula analyzes the artifacts less deeply and thus
    it builds a less complete model;
it also validates the model less thoroughly, so the model records fewer of the
    code-level behaviors.
With Haiku-4.5, the workload outstrips the LLM: \specula frequently stops,
    declaring the task done before finishing it.

\para{Cost.}
With Sonnet-4.6, \specula takes
    nearly the same time as with Opus-4.8 (\minssonnet{} vs.\ \minsopus{} minutes per
    module on average), and spends \budgetsonnet{} of the budget, yet finds $\frac{1}{6}$
    of the bugs: about \costpergtsonnet{} per bug found, against about
    \costpergtopus{} with Opus-4.8.
With Haiku-4.5, \specula costs $\frac{1}{17}$ as much per module as with Opus-4.8, but finds
    nothing.

\para{False positives.}
A weaker LLM visibly deviates from \specula's instructions.
With Sonnet-4.6, \fpcandsonnet{} false bugs reached the
    reproduction phase.
Among \fpcleansonnet{} of them, \specula followed the instructions: it attempted the reproduction
    as instructed, watched it fail, and eventually realized that they were not true bugs.
On the other six, it hacked the reproduction to make it succeed by injecting
    illegal states directly into the running system.
Such reward hacking is explicitly disallowed by the instructions.
With the same instructions, \specula with Opus-4.8 exhibits no such violation.

\section{Discussion}

\para{Durable design.}
The development of \specula has coincided with rapid advancement of LLM
capabilities. Earlier LLMs and agents often failed to produce high-quality \tla{} 
    specifications without substantial repair effort,
    while recent LLMs can carry much more of the modeling work on their own. 
This shift shaped our view of \specula's design---the system should not be limited by 
    specific LLM/agent generations, 
    but should improve as the LLMs/agents improve. 
The durable core of \specula is to treat system artifacts as the ground truth.
The agents decide the invariants and behaviors to model from the system artifact
    and check them against code bi-directionally: 
    trace validation confirms the model admits code-level executions, 
    and model checking exposes states the model incorrectly allows. 
Model-level violations return to code, 
    where reproduction decides whether they are true bugs or not. 
This structure remains necessary as LLMs improve,
    because hallucination, overgeneralization, and reward hacking are unlikely
    to disappear. 

\para{Evolving with LLMs/agents.}
With the durable design, 
    the implementation of \specula carries optimizations we made 
    to leverage the capabilities of the current coding agents. 
\specula's skills and tools are equipped to
    compensate for specific weaknesses of LLMs or coding agents today,
    and we expect to evolve them when LLMs/agents improve. 
Take bug reproduction  (\S\ref{sec:findbugs}) as an example. 
Reproducing a bug at the code level is a durable design, 
    but how we let the agent achieve the reproduction is not. 
Asked to reproduce a violation, a current coding agent often hacks out 
    a success:  it injects a faulty state directly or
    calls internal functions out of context.
\specula therefore restricts reproduction to behaviors 
    an external client can trigger, and forbids reward hacking. 
With restrictions in place, frontier agents produce almost
    no false positives and follow the instructions,
    while a weak LLM may not (\S\ref{sec:eval-models}). 
As LLMs/agents become even more capable, the way \specula carries out bug 
    reproduction can further improve.

\para{Resolving issues iteratively.}
Several components in \specula produce artifacts whose correctness is hard
    to guarantee at the time they are generated, e.g., code instrumentation,
    \tla{} models, and invariants.
Instead of forcing each artifact to be correct in local context,
    \specula accepts imperfect ones and resolves issues iteratively
    through self-evolving loops.
For example,
\specula generates invariants, without establishing that
    they must hold for the implementation, at the time 
    it extracts invariants from the system repository (\S\ref{sec:inv}).
This is a deliberate design because
demanding a correct invariant upfront is overly difficult at the moment
    when the agents know little about the system implementation.
When an invariant does not hold, 
    resolving it in the loop forces an investigation that deepens the
    agent's understanding of the system (\S\ref{sec:conformance}).

\para{Formal guarantee.}
\specula carries a few common limitations of AI-driven autonomous techniques---it 
is hard to provide formal guarantees of the behaviors and results 
    of LLM-based agents.
For example, \specula reads and comprehends 
    the system artifacts (e.g., its software repository)
    to decide what to model;
    there is no formal guarantee that it models everything perfectly
    as the modeling is autonomously driven by the agents.
Similarly, the invariants \specula extracts may not capture
every property a system should hold. A model can therefore remain
inconsistent with the implementation in ways we did not catch, and a bug
that depends on such a gap can remain undetected.

\section{Related Work}


\para{Model checking.}
Model checking aims to exhaustively explore the state space of 
    abstractions of actual programs~\cite{ClarkeE81} 
    and has been shown to be effective
    for reasoning about real-world systems~\cite{newcombe2015tlaaws,tang2024sandtable,ouyang2025remix,Converos}.
For concurrent and distributed systems, \tla{} is commonly used to 
    check system design and implementation~\cite{yu99tlc,newcombe2015tlaaws,hackett2023understanding,howard2025ccf}.
Applying model checking to real-world systems must
    carefully balance the cost of writing and maintaining formal specifications, 
    state space, and model-code conformance.
Implementation-level model checking~\cite{yang2009modist,guo2011demeter,leesatapornwongsa2014samc,lukman2019flymc} avoids 
    modeling effort by exploring state space directly at the code level;
    however, with the enormous amount of detail in code behaviors, it is hard to bound the state space in a general way.
Specification-based approaches~\cite{tang2024sandtable,ouyang2025remix,Converos,howard2025ccf} 
    lift exploration to a formal model, making state exploration 
    more tractable but requiring human experts to write the model, 
    choose the action granularity, 
    and check conformance with code.
Recent work~\cite{ouyang2025remix,Converos,gu2022compositional} develops multiple system models 
    at different levels of abstraction
    to help model checking focus on target modules (e.g., the changed ones).
\specula{} follows this practice using an agentic approach.
Unlike prior work that relies on human experts to decide the levels of abstraction,
    \specula autonomously makes the tradeoff between utility and efficiency
    based on target invariants.

\para{Model-code conformance.}
There are two common ways of conformance checking:
    replaying models at the code level~\cite{tang2024sandtable,ouyang2025remix}
    and validating code-level traces against the model~\cite{tasiran2003using,pressler2018trace,davis2020extreme_modelling,cirstea2024trace,howard2025ccf}.
Early refinement-style validation requires strict state and action mappings between code and its model~\cite{pressler2018trace,davis2020extreme_modelling}.
Recent work~\cite{cirstea2024trace,howard2025ccf,Converos,hackett2026omnilink} 
    relaxes this requirement by using \tla{}'s nondeterminism 
    to infer missing states and compose actions to align granularity, e.g.,
    OmniLink~\cite{hackett2026omnilink} validates code-level traces without 
    requiring each API call to have a single observed linearization point.
\specula{} uses trace validation to align generated models with code across different action granularities 
    and abstraction levels. Compared with traditional approaches, it automates instrumentation and trace checking.
Moreover, \specula{} couples trace validation with model checking 
    to catch overly permissive model repair that trace validation alone would accept.

\para{AI-based specification generation.}
AI-based specification generation has mostly followed the deductive verification approach 
    for sequential programs, where specifications are program annotations such as preconditions, postconditions, and loop invariants~\cite{ma2024specgen,wen2024enchanting,yang2025sespec}.
Recent work also translates natural-language descriptions into formal 
    specifications~\cite{cao2025acl,cosler2023nl2spec}.
Model checking instead requires specifications that define a behavioral model and properties to check.
Recent work also uses agents to construct and repair formal models, e.g.,
Event-B Agent synthesizes and repairs Event-B models from requirements using formal-tool feedback~\cite{wang2026eventb}.
\specula{} extracts both invariants and a model of the system from the target's own artifacts.
    It uses coding agents' semantic understanding of the artifacts and 
    generates scenarios that reduce the state space to explore in model checking. 

\para{Formal-methods agents.}
Recent agents use formal-methods concepts or tools to guide bug finding.
FM-Agent uses a top-down approach to derive natural-language function specifications from callers and applies AI-based Hoare-style reasoning over code blocks to flag violations~\cite{ding2026fmagent}.
BMC-Agent performs per-function bounded model checking by translating inferred pre/postconditions into CBMC/Kani harnesses, then validates and refines counterexamples~\cite{sun2026agenticmc}.
\specula{} differs by moving the reasoning target from function-level source code to system-level behaviors.
It targets distributed and concurrent system bugs expressed as protocol- and code-level invariant violations, using self-evolving loops to repair model, invariant, and instrumentation errors.

Recent agents target auto-generation of formal proofs in Lean, Coq/Rocq, Dafny, and 
    Verus~\cite{liu2026numina,tu2026autorocq,banerjee2026dafnypro,yang2025autoverus,liu2026kverus}.
These agents use verifier feedback to refine a formal artifact until the verifier accepts it.
These works can provide stronger guarantees than bug finding, but are scoped to the behaviors captured by the formalization and proof obligations.
\specula{} makes a different tradeoff: it prioritizes autonomous, low-cost specification and discovery of complex distributed and concurrency bugs, where system-wide deductive verification remains challenging and limited by practical constraints. 

\section{Remarks}

Our goal of developing the \specula project is to expand the reach of
    formal methods to improve the correctness and reliability of system software,
    with the astonishing capabilities of today's AI agents.
A key principle is to exploit the generative capabilities of AI to lower the cost 
    of formal specifications,
    while addressing the new challenges introduced by AI for specification generation.
\specula shows that a fully autonomous formal-methods system
    can be built with careful design and implementation,
    and it is feasible to achieve a reliable specification technique 
    atop imperfect (and often unreliable) AI agents.
\specula is under active development (hence we have decided not to have a conclusion section).
We encourage everyone to try it out, for fun and profit!

\section*{Acknowledgements}

The \specula project started from a hackathon effort to compete in the
    GenAI-accelerated TLA+ Challenge in 2025.
We thank the TLA+ Foundation for organizing the challenge
    and recognizing our work.
We thank Markus Kuppe for continuously supporting the project
    and for the invaluable discussions.
We thank Neil Giridharan for helping us understand our results on
    Autobahn.
We thank everyone who gave us feedback and/or encouragement throughout the development
    of \specula, including
    Zhizhen (Cathy) Cai, Zhiting Zhu, Matthew Malcomson, Dmitry Kulagin, Claudia Cauli,
    Timo Lang, Igor Konnov, Chris Newcombe,
    and Darko Marinov.

\bibliographystyle{acm}
\bibliography{main}


\end{document}